\documentclass[runningheads]{llncs}

\usepackage{amsmath,amssymb}
\usepackage{graphicx}
\usepackage{verbatim}
\usepackage{placeins}

\newtheorem{assn}{Assumption}
\newtheorem{non-assn}{Non-Assumption}

\DeclareMathOperator{\Var}{Var}

\begin{document}
\title{Estimating group properties in online social networks with a classifier\thanks{The authors thank members of the Social Dynamics Laboratory and anonymous reviewers for their helpful suggestions. The authors were supported while this research was conducted by grants from the U.S. National Science Foundation (SES 1357488), the National Research Foundation of Korea (NRF-2016S1A3A2925033), the Minerva Initiative (FA9550-15-1-0162), and DARPA (NGS2). The funders had no role in study design, data collection and analysis, decision to publish or preparation of the manuscript.}}
\titlerunning{Estimating group properties}

\author{
	George Berry\inst{1}\orcidID{0000-0003-3898-2380} \and
    Antonio Sirianni\inst{1}\orcidID{0000-0002-7710-3513} \and
    Nathan High\inst{1} \and
    Agrippa Kellum\inst{1} \and
    Ingmar Weber\inst{2}\orcidID{0000-0003-4169-2579} \and
    Michael Macy\inst{1}
}

\authorrunning{Berry et al.}

\institute{
	Cornell University, Ithaca, New York, USA\\
    \email{\{geb97,ads334,nmh53,ask252,mwm14\}@cornell.edu} \and
    Qatar Computing Research Institute, Doha, Qatar\\
    \email{iweber@hbku.edu.qa}
}

\maketitle

\begin{abstract}
We consider the problem of obtaining unbiased estimates of group properties in social networks when using a classifier for node labels. Inference for this problem is complicated by two factors: the network is not known and must be crawled, and even high-performance classifiers provide biased estimates of group proportions. We propose and evaluate AdjustedWalk for addressing this problem. This is a three step procedure which entails: 1) walking the graph starting from an arbitrary node; 2) learning a classifier on the nodes in the walk; and 3) applying a post-hoc adjustment to classification labels. The walk step provides the information necessary to make inferences over the nodes and edges, while the adjustment step corrects for classifier bias in estimating group proportions. This process provides de-biased estimates at the cost of additional variance. We evaluate AdjustedWalk on four tasks: the proportion of nodes belonging to a minority group, the proportion of the minority group among high degree nodes, the proportion of within-group edges, and Coleman's homophily index. Simulated and empirical graphs show that this procedure performs well compared to optimal baselines in a variety of circumstances, while indicating that variance increases can be large for low-recall classifiers.

\end{abstract}

\section{Introduction}

When seeking to understand social interaction online, researchers are commonly faced with a paradox: online data is behaviorally rich but lacks even basic demographic annotation. The lack of demographic information frustrates seemingly straightforward questions: for instance, what is the gender breakdown on an online platform? Since many important social science questions require demographic data, classifiers are commonly used to predict node-level attributes such as gender \cite{rao_classifying_2009,ciot_gender_2013,liu_whats_2013,fang_relational_2015}, education \cite{ding_predicting_2016}, age \cite{nguyen__2013}, race \cite{mohammady_using_2014,messias_white_2017}, income \cite{malmi_you_2016,culotta_predicting_2016}, or political affiliation \cite{al_zamal_homophily_2012-1,barbera_less_2016}. However, classifiers introduce error which can bias estimates of group properties, from demographic distributions to cultural homophily.

\begin{figure*}[ht]
\makebox[\textwidth][c]{\includegraphics[width=\linewidth]{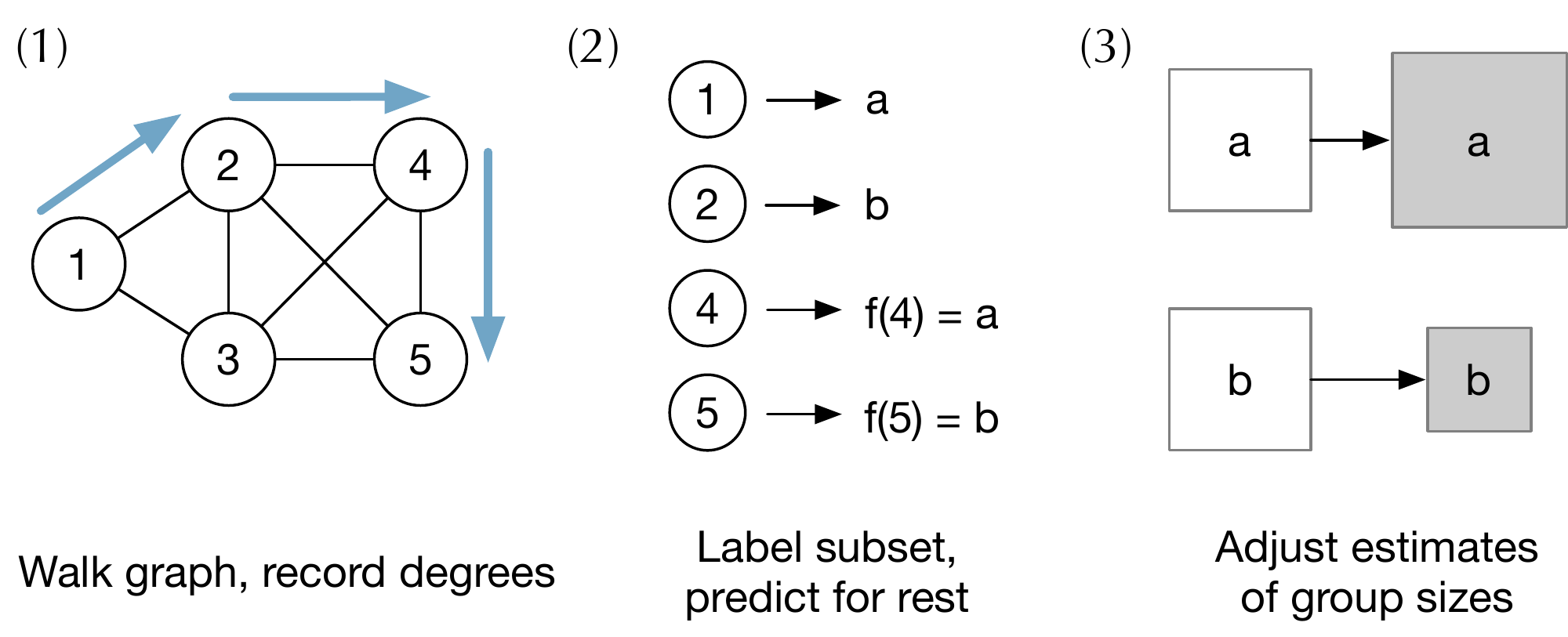}}
\centering
\caption{A demonstration of the steps in the AdjustedWalk process. First, the graph is randomly walked and node degrees are recorded. Second, a subset of the walked nodes are given ground-truth labels (nodes 1 and 2) and a machine learning model is used to predict for the remaining nodes in the sample (nodes 4 and 5). Third, the mean of the relevant quantity (e.g. proportion in $b$) is estimated using the RWRW estimator and then adjusted to remove bias.}
\label{fig:overview}
\end{figure*}

Adding to the challenge posed by limited demographic information, studies often rely on convenience samples of the underlying social network. The combination of classification error and non-representative sampling poses substantial challenges for obtaining valid estimates of group properties.

We propose a framework, which we term AdjustedWalk, to address these dual problems in order to obtain unbiased estimates of group properties in networks. The idea is to combine re-weighted random walk sampling (also called respondent driven sampling) \cite{gjoka_walk_2009,salganik_sampling_2004} with quantification learning \cite{forman_counting_2005,forman_quantifying_2008}. The sampling method provides the information necessary for inference over nodes and edges. A subset of the nodes is labeled and classified, and a post-hoc correction is applied to correct for classification bias at the group level.

We assume that the sampling procedure starts with an arbitrary node in an undirected graph, and that node labels are predicted by a possibly biased classifier with a known error rate. We show that the framework proposed performs well relative to baselines in four estimation tasks: group proportions, within-group edge proportions,  group visibility (the proportion of the minority group in the top 20\% of the degree distribution) \cite{karimi_visibility_2017}, and Coleman's homophily index \cite{coleman_relational_1958}.

The problem we study has three parts: drawing a representative sample from a network, building a classifier to predict node attributes, and correcting classification error to obtain unbiased estimates of group proportions. Many previous studies have examined each of these parts individually. Research on applications of respondent driven sampling (RDS) to online social networks has provided a robust toolkit for sampling online social networks when demographics are freely available from the site itself \cite{salganik_sampling_2004,goel_respondent-driven_2009,gjoka_walk_2009,gjoka_walking_2010,gjoka_practical_2011}. On the other hand, many studies have addressed the task of predicting demographics with high observation-level accuracy, with less attention paid to the representativeness of the sample or group-level estimates \cite{culotta_predicting_2016,fang_relational_2015,culotta_predicting_2015,volkova_inferring_2015,barbera_less_2016,malmi_you_2016,ciot_gender_2013,nguyen__2013,liu_whats_2013,ding_predicting_2016,gong_joint_2014,al_zamal_homophily_2012-1,wang_your_2016,messias_white_2017}. Finally, a literature on quantification learning \cite{forman_counting_2005,forman_quantifying_2008,gao_classification_2016} has addressed the problem of using a classifier to make population inferences. We demonstrate that combining these separate literatures allows social scientists to make better inferences about social groups in online social networks. This has applications for digital demography and can be useful for online-offline comparisons of social interaction, homophily, and representation \cite{karimi_visibility_2017}.

\subsection{Summary of contributions}

\begin{itemize}
\item This paper proposes AdjustedWalk, a framework for estimating group properties in online networks when neither the network nor group labels are known in advance. It contains the following steps: 1) a re-weighted random walk (RWRW) of the graph; 2) training a classifier by labeling a subset of the nodes walked; 3) adjusting group-level error introduced by the classifier to remove bias. The process is visualized in Figure \ref{fig:overview}. Importantly, walking the graph before labeling and classification makes it more likely that population-level inferences will be valid.
\item The performance of RWRW sampling is compared to three other plausible sampling procedures: node sampling, edge sampling, and snowball sampling \cite{wagner_sampling_2017}. RWRW performs well relative to the optimal sampling method for each task, and performs only slightly worse than node sampling overall despite the absence of a sampling frame.
\item An analytical expression for the increased variance of the adjusted estimate is provided, which can be expressed as a function of classifier recall.
\item AdjustedWalk is evaluated in a variety of conditions. We examine both simulated and empirical graphs, across a range of sampling fractions and classification accuracies. These analyses demonstrate that relatively small samples perform quite well. When considering classification accuracy, recall scores greater than 0.8 produce reasonable estimates, while recall scores lower than 0.8 quickly increase variance.
\item We discuss the conditions under which the results presented here apply to directed graphs in addition to undirected ones.
\end{itemize}

\section{Problem setup}

\subsection{Graph and groups}

We sample from a graph $G = (V, E)$, where $V$ indicates vertices and $E$ indicates edges. $N$ is the number of nodes in $G$. Call a node $i$ and an edge from $i$ to $j$ $e_{ij}$. Let $d_i$ be the degree of $i$, $D = \sum_i d_i$ the total degree of the graph, and $\bar{d} = D / N$ the average degree of the graph. We assume $G$ is undirected, so that $e_{ij} \in E \implies e_{ji} \in E$. We also assume there are no self links or multi-edges, that the graph has at least one triangle, that $G$ is connected, and that all edges have weight 1.

Nodes belong to one of two \emph{social groups}, denoted $a$ and $b$. By convention, $b$ is the minority group. These groups represent characteristics of individuals such as age, race, ethnicity, gender, or wealth status. $p_a$ is the proportion of nodes belonging to group $a$, and $\mathbf{p} = (p_a, p_b)$ is the vector of \emph{population proportions}. Since $p_a + p_b = 1$, we will frequently reason about one group with the implication that the same analysis holds for the other one. $s_{ab}$ is the proportion of edges from group $a$ to group $b$, with $\mathbf{s} = (s_{aa}, s_{ab}, s_{bb})$ the vector of \emph{edge proportions}. Since the graph is undirected, $s_{ab}$ represents all edges with one end in $a$ and the other in $b$.

When a classifier is used to categorize nodes, we do not observe $\mathbf{p}$ or $\mathbf{s}$ directly. We instead obtain estimates of these quantities after classification error. $\mathbf{m} = (m_a, m_b)$ is $\mathbf{p}$ after classification error, and $\mathbf{t} = (t_{aa}, t_{ab}, t_{bb})$ is $\mathbf{s}$ after classification error. We use hat notation (e.g.\ $\mathbf{\hat{m}}$) for estimates resulting from sampling part of the graph and then classifying the sampled nodes.

\subsection{Classification for quantification}

Assume we have a relationship between a set of features $x_i$ and an outcome $y_i \in \{a, b\}, y_i = f(x_i)$. A classifier approximates $f$, $y_i = \hat{f}(x_i) + \epsilon_i$. The model $\hat{f}$ makes classification errors, which are counted and stored in a confusion matrix

\begin{equation*}F = 
\begin{bmatrix}
\text{count}({\hat{a} \mid a}) & \text{count}({\hat{a} \mid b})\\
\text{count}({\hat{b} \mid a}) & \text{count}({\hat{b} \mid b})
\end{bmatrix}.
\end{equation*}

\noindent
We use the notation $(\hat{b} \mid a)$ to represent ``a true member of $a$ classified as a member $b$''. We will work with the column-stochastic \emph{misclassification matrix}, which we get by column-normalizing $\hat{f}$,

\begin{equation*}C = 
\begin{bmatrix}
c_{\hat{a} \mid a} & c_{\hat{a} \mid b}\\
c_{\hat{b} \mid a} & c_{\hat{b} \mid b}
\end{bmatrix},
\end{equation*}

\noindent
where $c_{\hat{b} \mid a} = \text{count}({\hat{b} \mid a})/\text{count}({\hat{a} \mid a}) + \text{count}({\hat{b} \mid a}))$ represents the probability of a true member of $a$ being classified as $b$. In practice $\hat{f}$ and $C$ are constructed via cross-validation and holdout sets.

The matrix $C$ is important for removing bias from estimates. We refer to the off-diagonal elements of $C$ as the ``misclassification rate'', and note that the diagonals are equivalent to both recall and accuracy. Precision depends on the relative sizes of groups. The misclassification rate represents the probability that a true member of group $a$ is classified as $b$, and vice versa.

If $C$ is known and only $\mathbf{m}$ is observed, $\mathbf{p}$ can be recovered. $C$ maps $\mathbf{p}$ to $\mathbf{m}$,

\begin{equation*}
\begin{bmatrix}
c_{\hat{a} \mid a} & c_{\hat{a} \mid b}\\
c_{\hat{b} \mid a} & c_{\hat{b} \mid b}
\end{bmatrix}
\begin{bmatrix}
p_a \\
p_b
\end{bmatrix}
=
\begin{bmatrix}
m_a \\
m_b
\end{bmatrix},
\end{equation*}

\noindent
which can be written compactly as

\begin{equation}
C \mathbf{p} = \mathbf{m}.
\end{equation}

\noindent
This implies that inverting $C$ provides a way to recover the true population proportions $\mathbf{p}$,

\begin{equation} \label{eq:p_hat}
\mathbf{p} = C^{-1} \mathbf{m}.
\end{equation}

This procedure is referred to as ``adjusted classify and count'' in the machine learning literature on \emph{quantification} \cite{forman_counting_2005,forman_quantifying_2008}, or recovering population proportions. In general, even high performance classifiers produce biased estimates of group proportions \cite{gao_classification_2016}, particularly for small groups. While group proportions may be both over- and under-estimated, classifiers tend to favor larger groups since loss functions are optimized at the observation level. This means that the size of large groups is often overstated by classifier predictions. Models which directly try to estimate group proportions \cite{gao_classification_2016} usually under-perform models trained at the observation level and then corrected as in Equation \ref{eq:p_hat}.

For quantification, an important assumption is required for inference to be valid from the individual to group level \cite{forman_quantifying_2008}.

\begin{assn} \label{assn:stable_features}
Stable conditional feature distribution
\begin{equation}
P_{\text{train}}(X = x_i | Y = y_i) = P_{\text{population}}(X = x_i | Y = y_i)
\end{equation}
\end{assn}

\noindent
Assumption \ref{assn:stable_features} states that the feature distribution within each class is the same in training set and population. This can fail for some (but not all) types of sample selection bias \cite{zadrozny_learning_2004,liu_robust_2014}. For instance, Assumption \ref{assn:stable_features} would fail to hold in a case where one group (e.g. men) tended to be sampled only if they had a certain feature (e.g. owned a car), and this feature was used in the model $\hat{f}$. In this case, car owners would be overrepresented in the training set relative to the population. Assumption \ref{assn:stable_features} allows sampling different groups at different rates, as long as the samples drawn from within each group preserve the within-group feature distribution.

If these assumptions hold, then $C_{\text{train}} = C_{\text{population}}$. This implies that bias can be corrected in the entire sample. If the sample is drawn from the population, this provides an inference about the population quantity.

We propose conducting the network walk \emph{before} labeling and classifying cases so that Assumption \ref{assn:stable_features} is more likely to be satisfied. In this case, if the walk draws a valid sample from the population, then Assumption \ref{assn:stable_features} holds. In cases where a classifier is trained on cases not from the sample, Assumption \ref{assn:stable_features} cannot be tested since $y_i$ is unknown in the sample. We also assume that the misclassification matrix $C_{train}$ learned from the labeled set is known exactly. In practice, there is some uncertainty around the confusion matrix, although this can be addressed both through cross validation and the use of holdout set.

\subsection{Matrix adjustment vs. calibration}

We work with the class labels, but an alternative to the matrix method discussed here is to train a second model which calibrates $\hat{f}$. A calibration model takes predicted scores $\hat{s}_i \in [0, 1]$ from $\hat{f}$ and fits a model so that $\hat{s}_i \approx P(y_i = a)$. Gao and Sebastiani \cite{gao_classification_2016} examine a variety of quantification methods on many natural language processing tasks and find that the ``adjusted classify and count'' method used in this paper is statistically indistinguishable from using a calibrated model. However, the calibrated model has somewhat better average performance.

When considering the increased variance of adjustment methods, a closed-form can be derived for the matrix method (shown in the Appendix). A calibration method has the potential to provide lower variance estimates because it incorporates more information. However, it can also be more difficult to assess the increased variance because of correlated errors between observations which enter into the variance. For simplicity, we study the matrix method and note that variance may be reduced by instead fitting a calibration model.

\subsection{Graph walking} \label{sec:rds}

Re-weighted random walking (RWRW) is a Markov Chain Monte Carlo (MCMC) sampling procedure \cite{goel_respondent-driven_2009}. It is also commonly referred to as Respondent Driven Sampling (RDS) \cite{salganik_sampling_2004}, which is a process for applying RWRW to offline social networks such as those of jazz musicians \cite{heckathorn_finding_2001} or injection drug users \cite{ramirez-valles_networks_2005}.

RWRW allows randomly walking a connected, undirected graph and obtaining a valid estimate of node properties through reweighting by node degree. The basic intuition is to conduct a random walk of the undirected graph $G$, recording node degrees along the walk. The walk itself provides a random sample of edges, while degree information can be used to approximate a random sample of the nodes.

We present only the RWRW estimator here. A derivation may be found in \cite{goel_respondent-driven_2009}. A more general introduction to the method may be found in \cite{salganik_sampling_2004,volz_probability_2008}. For extensions to online network applications, see \cite{ribeiro_estimating_2010,kurant_walking_2011}.

Assume we wish to take a mean of a function $g$ over the nodes $i$ of graph $G$, where $g$ is an indicator function for $i$ being a member of minority group $b$. Choose a seed node with probability $\pi(i) = d_i / D$, or proportional to the node's degree. Then randomly walk the graph starting from the seed for $n$ steps. Each jump samples node $i$ with probability $\pi(i) = d_i / D$. For each node $j$ along the walk, we record the node's degree $d_j$ and $g(X_j)$, where $X_j$ represents the $j$th node encountered while walking. The RWRW estimator for estimating the proportion of nodes belonging to $b$ (assume no classification error) is given by

\begin{equation}\label{eq:rds_estimator}
\hat{p}_b = \frac{1}{\sum_{j=0}^{n-1} 1 / d_j} \sum_{j=0}^{n-1} \frac{g(X_j)}{d_j}.
\end{equation}

\noindent
Since the random walk naturally samples nodes proportional to degree, we correct for this through the weighting procedure in Equation~\ref{eq:rds_estimator}. This estimator provides an asymptotically unbiased estimate of $p_b$.

Note that $g$ may be a continuous valued function as well, allowing estimates of mean degree and the degree distribution (see \cite{ribeiro_estimating_2010} for an example).

In addition to a mean of $g$ over the nodes of the graph, the RWRW process records information which may be used to estimate the distribution of $g$. We discuss the details in the Appendix, and use this fact to estimate quantiles of the degree distribution for estimating node visibility.

\subsection{Walking in a directed graph}

The RWRW procedure relies on an important property of walks in undirected graphs: the probability of being at a given node at any time is proportional to that node's degree. However, many online networks are directed, such as Twitter follow relationships. If the directed graph can be redefined as an undirected one, than a walk can still be conducted. For instance, if both inlinks and outlinks can be accessed when a node is visited, then the combined inlink-outlink graph can be walked as if it were undirected. If this is not possible, additional approaches are possible but beyond the scope of this paper (see \cite{gjoka_practical_2011} for suggestions).

\section{Results}

We study four outcomes: proportion of nodes in the minority group, the fraction of in-group edges in the minority group, the percentage of minority group members in the top 20\% of the degree distribution (visibility), and Coleman's homophily index.

Node and edge proportions are straightforward measures. Visibility has been studied in recent work \cite{karimi_visibility_2017,wagner_sampling_2017}. It can indicate lack of status among minority group members. For instance, if the minority group comprises 20\% of the population but only 10\% of the top quintile of the degree distribution, minority group members are systematically underrepresented in the highest status positions.

Coleman's homophily index \cite{coleman_relational_1958} has long been employed by social scientists. It treats random mixing as a baseline, with a value of 1 indicating perfect homophily and -1 indicating perfect heterophily. For group $a$, this measure is defined as

\begin{equation} \label{eq:homophily}
H_a = \left\{\begin{array}{lr}
        \frac{s_a - p_a}{1 - p_a} & \text{for } s_a - p_a \ge 0\\
        \frac{s_a - p_a}{p_a} & \text{for } s_a - p_a < 0
        \end{array}\right\}.
\end{equation}

\subsection{Simulation parameters}

We follow the procedure in \cite{wagner_sampling_2017} to generate homophilous power-law graphs for the simple case of two groups, $a$ and $b$. Graphs have 10,000 nodes, mean degree of $8$, minority group fraction of 0.2, ingroup preference parameter of 0.8 (strong ingroup preference). The graph generation procedure balances ingroup preference with preference for links to high degree nodes. Note that the preference parameter is not the Coleman homophily index, but is correlated with the homophily index.

These graph parameters present a challenge for the technique presented in this paper, since RWRW is known to have higher variance in homophilous networks.\footnote{We also conducted simulations with minority group sizes of 0.35 and 0.5, and ingroup preferences of 0.2 (heterophily) and 0.5. The case we present is on balance the most challenging, although heterophilous graphs can present difficulties as well. We omit these additional cases for brevity, and because homophilous graphs are the case we are most often faced with empirically.}

Misclassification rates (off-diagonals of $C$) considered are 0\%, 10\%, 20\% and 30\%. We discuss the 20\% case most frequently because it is high enough to present a challenge for our method and low enough to correspond to much published research. We draw samples of size 1000 to 3000, in increments of 500.

We compare the proposed RWRW method with three other sampling methods that have been used in recent literature: random node sampling, random edge sampling, and snowball sampling \cite{wagner_sampling_2017}. These methods provide reasonable comparisons: we expect node sampling to outperform RWRW for node-level tasks and edge sampling to outperform RWRW for edge-level tasks. Snowball sampling is included since it is a convenience sampling method that was commonly used before RWRW was developed.

\subsection{Group proportions and visibility}

\begin{figure*}[ht]
\makebox[\textwidth][c]{\includegraphics[width=.95\linewidth]{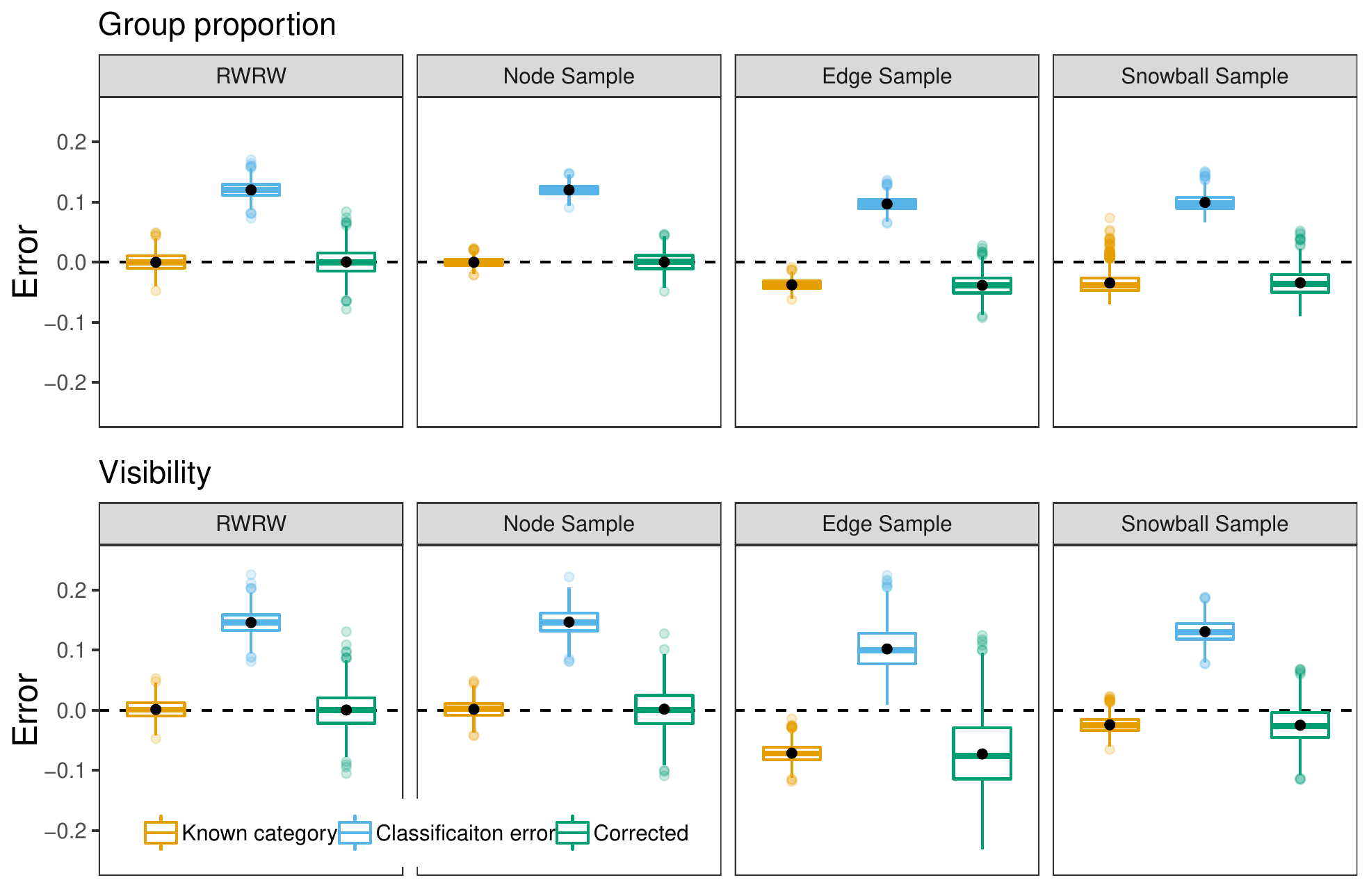}}
\caption{Estimates of group proportions and minority group visibility across sampling methods for 20\% classification error and sample size of 3000. For each plot, estimates are presented for three cases: no classification error, classification error, and corrected classification error. The dashed line at 0 indicates no error, and the black dot for each estimate indicates the mean. Whiskers represent 95\% of the distribution. RWRW performs well in both cases, while node sampling has the lowest variance as expected. Correcting for classification error removes bias only for RWRW and node sampling, while edge sampling and snowball sampling demonstrate bias even after correction.}
\label{fig:nodes_visibility}
\end{figure*}

Figure~\ref{fig:nodes_visibility} shows node proportion and visibility estimates across the four sampling methods for a 20\% misclassification rate and a sample size of 3000\footnote{The process for estimating the degree distribution for visibility is described in the Appendix.}. Performance for each sampling method is presented for three cases: no classification error, classification error with no correction, and classification error with correction. We present error distributions rather than the standard normalized root mean squared error \cite{ribeiro_estimating_2010,kurant_walking_2011} in order to assess bias.

\begin{figure*}[ht]
\makebox[\textwidth][c]{\includegraphics[width=.95\linewidth]{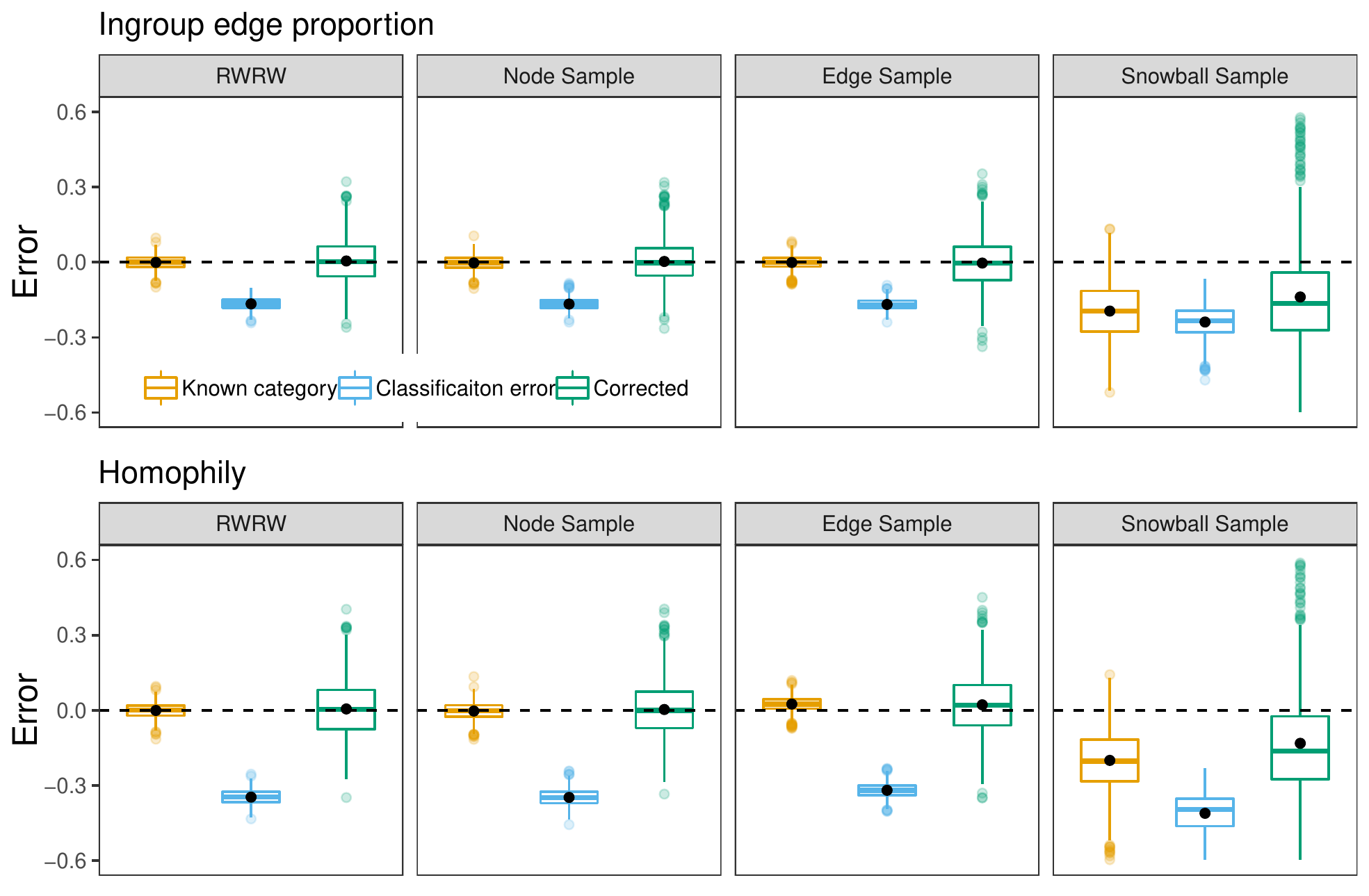}}
\centering
\caption{Estimates of ingroup edge proportions and group homophily across sampling methods for 20\% classification error and sample size of 3000. For each plot, estimates are presented for three cases: no classification error, classification error, and corrected classification error. The dashed line at 0 indicates the true value, and the black dot for each estimate indicates the mean. Whiskers represent 95\% of the distribution. Ingroup edge proportions are sampled well by RWRW, node sampling, and edge sampling. Homophily is more difficult but is well captured by both RWRW and node sampling.}
\label{fig:edges_homophily}
\end{figure*}

\begin{figure}[ht]
\includegraphics[width=.95\linewidth]{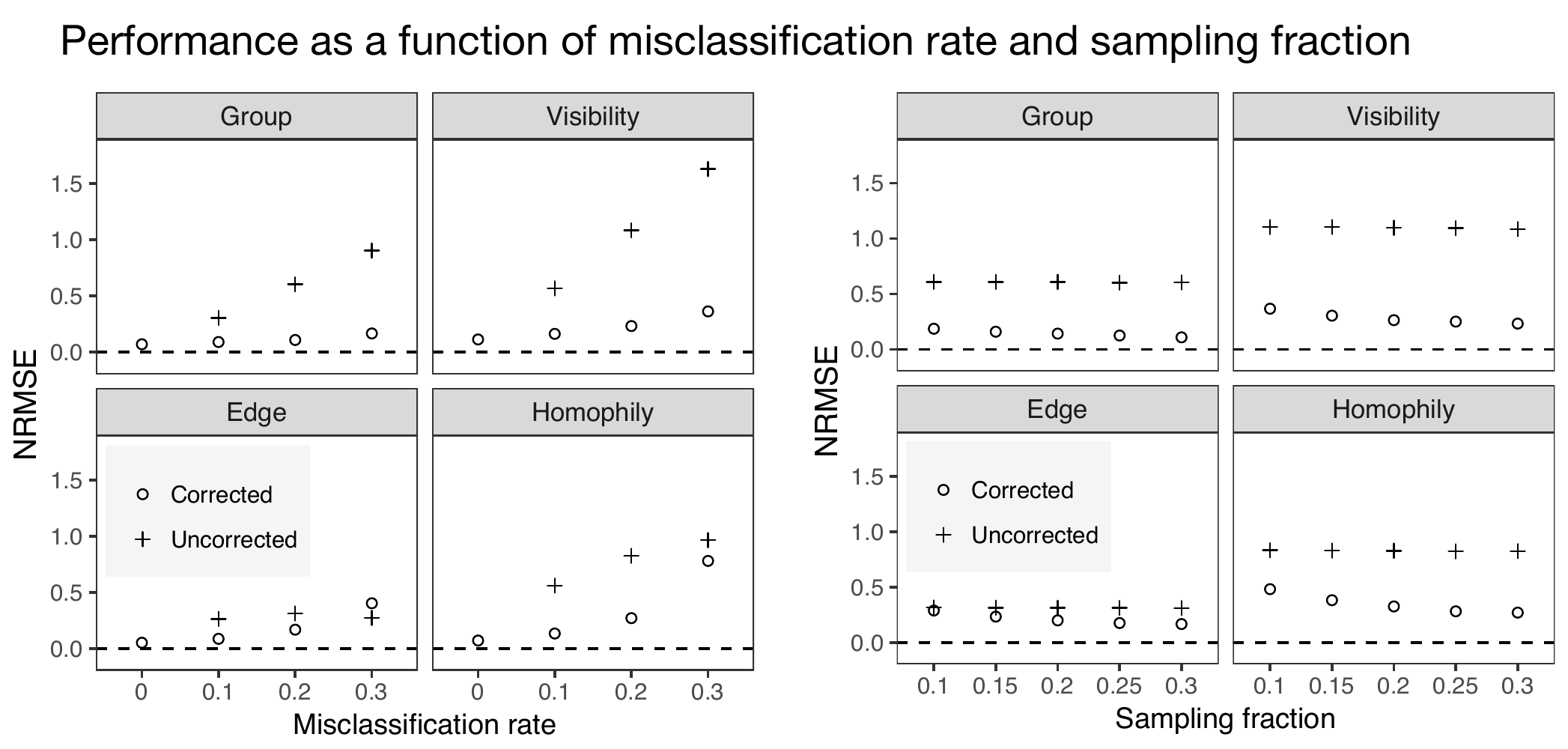}
\centering
\caption{Performance of RWRW as a function of misclassification rate and sampling fraction. For misclassification rate, a fixed walk size of 3000 is chosen. for a sample size of 3000. Circles indicate that bias correction has been applied, while plus signs indicate no correction. For group proportions and visibility, bias correction reduces NRMSE by a large amount. For edge proportions, the NRMSE reduction is smaller and disappears completely for misclassification rates of 0.3. Homophily shows large gains at misclassification rates of 0.1 and 0.2, and smaller gains at 0.3. In all cases, increasing sample size produces better after-correction estimates while barely changing before-correction estimates.}
\label{fig:rds_at_misclasf}
\end{figure}

For group proportions, node sampling and RWRW are both unbiased, and node sampling has a lower variance as expected. Edge and snowball sampling perform poorly. For visibility, RWRW performs the best since it does not have bias in the corrected case, while applying the correction to node sampling does not remove bias.

A 20\% misclassification rate without correction introduces a large magnitude error. While $p_b$ is 20\% in the population, misclassification causes this to rise to around 30\% in all cases, an inflation of 50\%. Since uncorrected estimates are low variance, a draw from an uncorrected distribution is likely to produce a poor estimate of the true value. In cases where assessing the size of a minority group is of importance, the upwardly biased estimates can lead to conclusions that understate the differences between groups. Smaller groups are likely to incur larger relative upward biases, since classifier loss functions penalize errors at the observation level and therefore tend to make errors at greater rates on smaller groups.

\subsection{Edge proportions and homophily}

Figure~\ref{fig:edges_homophily} shows error for estimates for edge proportions and homophily. The variance for these measures is larger than for the node-level measures (note the rescaled axes compared to Figure~\ref{fig:nodes_visibility}).

Homophily in particular is a difficult inference task since it combines estimates of both group sizes and interaction rates. The error magnitude for uncorrected homophily estimates is large, averaging -0.35 on a -1 to 1 scale. Crossing 0 for the homophily score causes a qualitative difference in interpretation. This error can easily turn an insular minority into a gregarious plurality. In the graphs studied here, the true average homophily value is 0.42, meaning that the average error introduced by misclassification is about 83\% of the true value.

The corrected estimates remove bias in exchange for an increase in estimate variance. For this level of misclassification (20\%), we argue researchers should take the variance in exchange for bias reduction. After correcting for classification error, RWRW produces estimates with error drawn from $\mathcal{N}(0.005, 0.136)$ while the 95th percentile of the uncorrected error distribution is -0.282 (this is the ``top'' closest to the no-error line). This indicates that nearly all draws (about 98\%) from the corrected distribution are lower-error than even the best draws from the uncorrected one.
\begin{figure}[ht]
\includegraphics[width=0.9\linewidth]{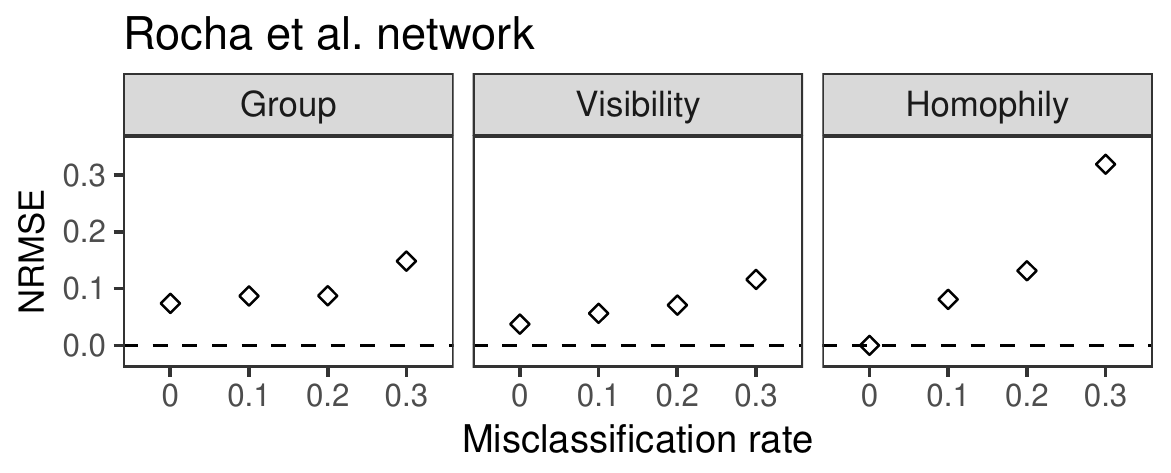}
\centering
\caption{Performance of RWRW after bias correction on the sexual contact network in \cite{rocha_simulated_2011} as misclassification rate is increased, with sample size given in Table \ref{tab:summary_stats}. The denominator of the NRMSE is 0 for edge proportions, so we omit that measure. Akin to other studied graphs, this network shows reasonable performance at low misclassification rates, although group proportion NRMSE is somewhat higher than expected. This is potentially caused by the perfectly heterophilous nature of the graph. At high misclassification rates, NRMSE becomes quite large for homophily and begins to rapidly increase for other measures.}
\label{fig:sc_network}
\end{figure}

\subsection{Sample size and misclassification rates}

Both increasing sample size and building better classifiers have costs. Figure \ref{fig:rds_at_misclasf} reports the sensitivity of RWRW estimates to sample sizes and classification error rates. Corrected RWRW estimates are unbiased, and are compared to uncorrected estimates, which are biased but have lower variance. The metric of performance used is the normalized root mean squared error (NRMSE), given by

\begin{equation*}
NRMSE = \frac{\sqrt{E[(\hat{\theta} - \theta)^2]}}{\theta}.
\end{equation*}

When increasing classification error (Figure~\ref{fig:rds_at_misclasf}, first panel), two patterns emerge. For group proportions and visibility, bias correction substantially reduces NRMSE at all misclassification rates. On the other hand, the NRMSE reduction is modest for edge proportions at misclassification rates of 0.1 and 0.2, while the uncorrected estimate has lower NRMSE at a misclassification rate of 0.3. For homophily, there is substantial reduction in NRMSE at misclassification rates of 0.1 and 0.2, with a more modest correction at misclassification rate of 0.3.

When examining NRMSE response to sample size (Figure~\ref{fig:rds_at_misclasf}, second panel) while holding misclassification at 20\%, we find an interesting pattern: increasing sampling fraction has little effect on the NRMSE of the uncorrected estimate for all measures, but reduces NRMSE for the corrected estimates. This indicates that increasing sampling fraction without correcting for classifier error may not bring estimates closer to the truth. In the simulated graphs here, tripling the sample size hardly changes the error without correction.

\section{Empirical graphs}

We study two empirical graphs, the social network Pokec \cite{takac_data_2012} and the sexual contact network studied in \cite{rocha_simulated_2011}. The Pokec graph \cite{takac_data_2012} was crawled from an online social network in Slovakia and comes labeled with gender and age attributes. Age was binarized into ``over 28 years old'' or not, corresponding roughly to the top quintile of the distribution. The sexual contact network \cite{rocha_simulated_2011} was collected in Brazil and represents links between high end escorts and their clients. We study gender as our category of interest, and note that this network is perfectly heterophilous (all ties are between men and women). Summary statistics for both networks can be found in Table~\ref{tab:summary_stats}. These graphs were preprocessed by limiting the graph to mutual ties in the largest connected component, and then deleting nodes which had missing values for the relevant demographic. In the case of the Pokec graph, where two different demographics are examined, we created separate graphs for each demographic.

\begin{table}
\centering
\begin{tabular}{ c c c c }
                           & Pokec gender & Pokec age         & Sexual contact \\ 
                           \hline
 Number of nodes           & 1,198,235    & 764,845           & 15,810         \\  
 Number of edges           & 8,312,749    & 4,172,385         & 38,540         \\
 Group studied             & Female       & $>$ 28 years old  & Female         \\
 Group proportion          & 51.3\%       & 22.2\%            & 39\%           \\
 Group homophily           & 0.034        & 0.344             & -1             \\
 Sample size               & 25,000       & 25,000            & 3,000\\
 \hline
\end{tabular}
\caption{Summary statistics for empirical graphs}
\label{tab:summary_stats}
\end{table}

\begin{figure*}[ht]
\makebox[\textwidth][c]{\includegraphics[width=.95\linewidth]{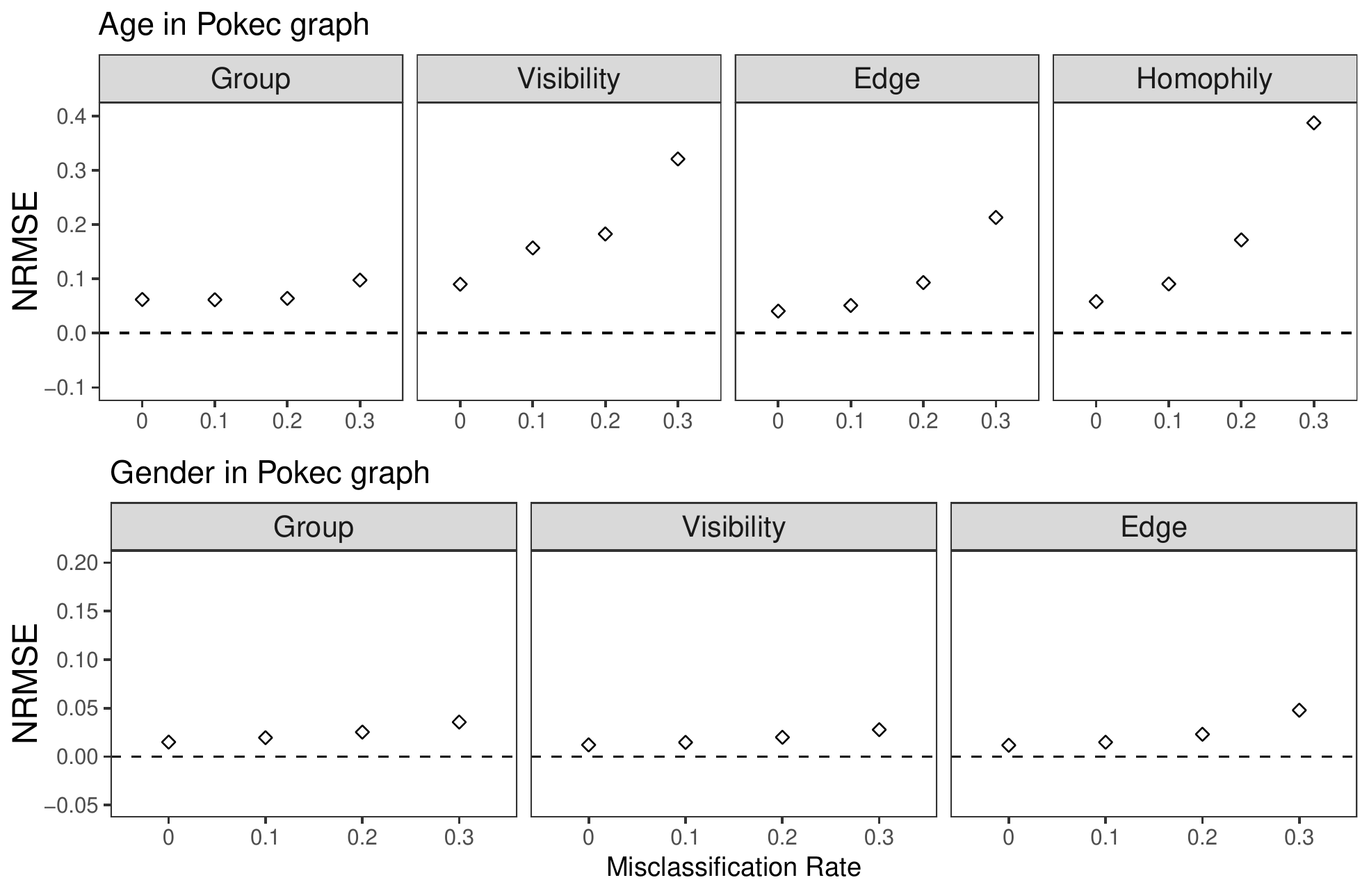}}
\centering
\caption{RWRW performance after bias correction on two graphs derived from the Pokec network \cite{takac_data_2012,snapnets}, with sample size given in Table \ref{tab:summary_stats}. The top panel presents results for age, which is moderately homophilous. The pattern of quickly accelerating error at misclassification rates of 0.3 appears again. The bottom panel shows performance on gender in the Pokec graph, where we omit the homophily measure since its true value is very close to 0 which inflates NRMSE. This is the homophily value closest to 0 that we study in the paper, and as expected performance is quite good (note the y-axis on the bottom panel). Ingroup edge proportion is typically difficult to sample, but even with a misclassification rate of 0.3, performance is strong. This suggests that when graphs do not display large amounts of homophily, less accurate classifiers can be used or smaller samples drawn.  }
\label{fig:pokec_network}
\end{figure*}

Figures \ref{fig:sc_network} and \ref{fig:pokec_network} present results for these empirical graphs using RWRW and bias correction as misclassification rate is increased. Performance is comparable to the simulated graphs. For the low-homophily graph (Pokec gender), sampling is quite easy at all misclassification levels. The sexual contact network presents a possible challenge since it is perfectly heterophilous, but performance is in line with the other graphs considered. The Pokec age graph has a moderate amount of homophily, although sampling a relatively small proportion of the network (25,000 out of 764,000 nodes) produces reasonable results. 

\section{Limitations} \label{sec:limitations}

There are several limitations to the work presented here, which researchers should consider carefully. Most importantly, we have assumed that the classifier error rate is known. Cross validation and holdout sets can be used to estimate this quantity, but empirically some uncertainty will still exist. An alternate route may be to train a calibrated model, although then the challenge becomes estimating the calibration accuracy. In some cases, either with small training sets or difficult classification problems, there may be substantial uncertainty about classifier performance.

Beyond this issue, online networks present data quality issues which may impact the validity of estimates. For instance, bots or inactive accounts which appear to belong to a certain demographic group can bias estimates. Accounting for bias from these factors can be difficult.

\section{Conclusion}

Classifiers offer an opportunity for social scientists to study demographics and other attributes online, in combination with rich behavioral data often absent from offline surveys. However, care must be taken in order for classification to yield accurate population-level estimates, particularly in networked online settings where the sampling frame is unknown. The framework we study in this paper provides a method to obtain unbiased estimates of group proportions in this setting. Additionally, it can be applied in many domains and requires relatively small adjustments to existing practice. This offers new opportunities for digital demography and comparisons of online settings with offline survey data.

\section{Appendix}

\subsection{Variance of corrected estimates}

Consider the simple case of classifying group proportions with two groups. We obtain a sample from the population with true proportions $\hat{p}$ and estimated proportions $\hat{m}$. Multiplying out Equation \ref{eq:p_hat} gives expressions for for the estimated $\hat{p}_a$ and $\hat{p}_b$,

\begin{equation}
\hat{p}_a = \frac{\hat{m}_a c_{\hat{b} \mid b} - \hat{m}_b c_{\hat{a} \mid b}}{\det(C)}, \quad
\hat{p}_b = \frac{\hat{m}_b c_{\hat{a} \mid a} - \hat{m}_a c_{\hat{b} \mid a}}{\det(C)}.
\end{equation}

The variance of the mean is given by,

\begin{align}
\Var(E[\hat{p}_a]) &= \Var(\frac{E[\hat{m}_a] - c_{\hat{a} \mid b}}{\det(C)}) \\
                &= \frac{1}{\det(C)^2}\Var(E[\hat{m}_a]),
\end{align}

\noindent
where we use the assumption that $C$ is constant to pull it out of the variance expression.

When there is no classification error, $\det(C) = 1$, and when the classifier guesses randomly (.5 in every cell), $\det(C) = 0$ and the variance is undefined. $\det^2(C)$ provides a clear quantification of the variance increase we expect for group proportions. For instance, if $C = [0.8, 0.2; 0.2, 0.8]$ with $\det^2(C) = 0.6$, we expect a variance increase of $1/0.6^2 = 2.78$. If classifier performance improves to $C = [0.9, 0.1; 0.1, 0.9]$, the variance increase is $1/0.8^2 = 1.56$.

$\Var(E[\hat{m}_a])$ comes from the random walking procedure itself and is generally not known in closed form. Two methods for closed-form variance have been proposed \cite{volz_probability_2008,goel_respondent-driven_2009}. The Volz-Heckathorn \cite{volz_probability_2008} estimator is biased but provides reasonable estimates in practice. The Goel-Salganik \cite{goel_respondent-driven_2009} variance estimator relies on knowing the homophily of the network. Bootstrap resampling methods based on creating ``synthetic chains" from the estimated transition matrix between groups have also often been used \cite{heckathorn_respondent-driven_2002}.

Simulations of various RWRW estimators \cite{gile_respondent-driven_2010} show that factors such as a non-equilibrium seed selection, group homophily, and number of waves from each seed affect both the bias and variance of RWRW estimates. Generally, one long chain provides the best results, rather than many shorter chains. It is easier to sample from lower homophily networks, and equilibrium seed selection (proportional to degree) is useful if one must use relatively short chains. Otherwise, if chains may be long, a burn-in period can be used to simulate equilibrium seed selection.

\subsection{Correcting visibility}

While RWRW gives the mean of $g$ over the population of nodes, the distribution of $g$ is often an object of interest. For instance, if we wish to estimate the proportion of minority group members in the top 20\% of the degree distribution, we need to estimate the joint distribution of $(g(i), d_i)$ and take nodes in the top 20\% of the distribution of $d_i$.

Fortunately, importance resampling \cite{rubin_calculation_1987,mcallister_bayesian_1997} based on the data obtained during an RWRW walk provides a method to do this. If we know node $i$ with degree $d_i$ is sampled with probability $\pi(i)$ and we want to sample it with probability $1/N$ (a uniform distribution over the nodes), then we construct an importance weight using the ratio of desired over actual distributions

\begin{equation} \label{eq:importance_weight}
\frac{1/N}{\pi(i)} = \frac{D}{N d_i} = \frac{\bar{d}}{d_i} = w_i.
\end{equation}

\noindent
$w_i$ provides a resampling weight for node $i$. We then normalize $w_i / \sum_j w_j$ and resample data $(f(i), d_i)$ according to this probability to approximate draws from the desired distribution $1/N$.

An importance resample produces a distribution of $(d_i, g(i))$ which mirrors the distribution in the population. We then sort the resampled nodes by degree $d_i$ and take the proportion in the top 20\% of degree where $g(i) = b$, or where $i$ is a member of the minority group. In the case with no classification error, this procedure produces an unbiased estimate  of the fraction of minority group members in the top 20\% of the degree distribution.

With classification error, we need to add an additional step to correct the importance resample. Call $\hat{m}_b^{\mathcal{I}}(20)$ the measured proportion of group $b$ in the top $20\%$ of the degree distribution in importance resample $\mathcal{I}$. Likewise, there is a vector that contains measures for all groups $\mathbf{\hat{m}}^{\mathcal{I}}(20)$. Then we can use a procedure similar to Equation~\ref{eq:p_hat} to correct the importance resample proportions: 

\begin{equation} \label{eq:importance_resample}
\mathbf{\hat{p}}^{\mathcal{I}}(20) = C^{-1}\mathbf{\hat{m}}^{\mathcal{I}}(20).
\end{equation}

To see when $\mathbf{\hat{p}}^{\mathcal{I}}(20)$ is unbiased, repeat the reasoning for estimating the population proportion $\mathbf{\hat{p}}$ above. This shows that $\mathbf{\hat{p}}^{\mathcal{I}}(20)$ is unbiased when the importance resample provides an unbiased estimate of $\mathbf{m}^{\mathcal{I}}(20)$. A similar argument applies for the variance, and the determinant of $C$ may be used to estimate the increase in variance.

\subsection{Correcting edge proportions} \label{sec:correction_edge_prop}

Correcting estimates of ties between groups presents a more substantial challenge than correcting group proportions. Akin to $C$, there is a dyadic misclassification matrix $M$ which maps $\mathbf{s}$ to $\mathbf{t}$,

\begin{equation}
M \mathbf{s} = \mathbf{t},
\end{equation}

\noindent
where

\begin{equation*}
M = 
\begin{bmatrix}
c_{\hat{a} \mid a}^2 & c_{\hat{a} \mid a} c_{\hat{a} \mid b} & c_{\hat{a} \mid b}^2\\
2 * c_{\hat{a} \mid a} c_{\hat{b} \mid a} & c_{\hat{a} \mid a} c_{\hat{b} \mid b} + c_{\hat{a} \mid b} c_{\hat{b} \mid a} & 2 * c_{\hat{a} \mid b}  c_{\hat{b} \mid b} \\
c_{\hat{b} \mid a}^2 & c_{\hat{b} \mid a} c_{\hat{b} \mid b} & c_{\hat{b} \mid b}^2
\end{bmatrix},
\end{equation*}

\noindent
which implies that we can use a technique similar to Equation~\ref{eq:p_hat} at the dyad level

\begin{equation}
\mathbf{s} = M^{-1} \mathbf{t}.
\end{equation}

In practice, we obtain a sample $\mathbf{\hat{t}}$ rather than $\mathbf{t}$ for the entire graph, which is then used to estimate true edge proportions $\mathbf{\hat{s}}$. $\mathbf{\hat{s}}$ is unbiased when the sampling method employed produces unbiased estimates of $\mathbf{t}$. If $B = M^{-1}$, the expectation for $\hat{s}_a$ is given by

\begin{equation}
E[\hat{s}_{aa}] = b_{00}E[\hat{t}_{aa}] + b_{01}E[\hat{t}_{ab}] + b_{02} E[\hat{t}_{bb}].
\end{equation}

As in the node case, we can expect the variance of $E[\hat{s}_{aa}]$ and $E[\hat{s}_a]$ to increase when applying classification bias correction. Simulations below indicate that variance inflation for $E[\hat{s}_a]$ is larger than for $E[\hat{p}_a]$. Note that $\hat{s}_a = 2\hat{s}_{aa} / (2\hat{s}_{aa} + \hat{s}_{ab})$ is unbiased under the same conditions as $\hat{s}_{aa}$.

If $B = M^{-1}$, then the variance for $\hat{s}_{aa}$ is

\begin{equation}
\Var(E[\hat{s}_{aa}]) = b^2_{00}\Var(E[\hat{t}_{aa}]) + b^2_{01}\Var(E[\hat{t}_{ab}]) + b^2_{02} \Var(E[\hat{t}_{bb}]).
\end{equation}

\bibliographystyle{splncs04}
\bibliography{library}

\end{document}